\documentclass[11pt,oneside,onecolumn,draftclsnofoot,a4paper]{IEEEtran}
\usepackage{amsmath}
\usepackage{amssymb}
 \usepackage[mathscr]{eucal}

\usepackage[dvipsnames,usenames]{color}
\usepackage[dvips]{graphicx}
\usepackage{epsfig}
\usepackage[hang,raggedright]{subfigure}

\usepackage{soul}
  \usepackage[usenames,dvipsnames]{color}
  \usepackage{color}

\definecolor{lightblue}{rgb}{.5,.85,1}
\definecolor{lightred}{rgb}{1,.6,.5}
\definecolor{lightorange}{rgb}{1,.7,.4}

\def\comment#1{}
\def\bigcircledast{\mathop{\mbox{\fontsize{18}{19}\selectfont $\circledast$}}}

\def\diag{\operatorname{diag}}
\def\dvec{\operatorname{dvec}}
\def\vectorize{\operatorname{vec}}
\newcommand{\vtr}[1]{\vectorize\hspace{-.3ex}\left(#1\right)}

\newcommand{\ba}{{\bf a}}
\newcommand{\bS}{{\bf S}}
\newcommand{\bR}{{\bf R}}
\newcommand{\bD}{{\bf D}}
\newcommand{\bF}{{\bf F}}
\newcommand{\bY}{{\bf Y}}

\newcommand{\bW}{{\bf W}}
\newcommand{\bV}{{\bf V}}
\newcommand{\bM}{{\bf M}}
\newcommand{\bA}{{\bf A}}
\newcommand{\bK}{{\bf K}}
\newcommand{\bG}{{\bf G}}
\newcommand{\bI}{{\bf I}}\newcommand{\br}{{\bf r}}

\newcommand{\bZ}{{\bf Z}}
\newcommand{\bP}{{\bf P}}
\newcommand{\bH}{{\bf H}}
\newcommand{\bX}{{\bf X}}
\newcommand{\bE}{{\bf E}}
\newcommand{\bg}{{\bf g}}
\newcommand{\bs}{{\bf s}}\newcommand{\bff}{{\bf f}}
\newcommand{\bb}{{\bf b}}\newcommand{\bc}{{\bf c}}\newcommand{\bd}{{\bf d}}
\newcommand{\ve}{{\bf e}}

\newcommand{\I}{{\bf I}}

\newcommand{\bB}{{\bf B}}
\newcommand{\bQ}{{\bf Q}}

\newcommand{\bJ}{{\bf J}}
\newcommand{\bC}{{\bf C}}
\newcommand {\thetab} {\mbox{\boldmath{$\theta$}}}
\newcommand {\bPsi} {\mbox{\boldmath{$\Psi$}}}
\newcommand {\bPhi} {\mbox{\boldmath{$\Phi$}}}
\newcommand {\Zetab} {\mbox{\boldmath{$\Xi$}}}

\newcommand{\bGamma}{\mbox{\boldmath $\Gamma$}}
\newcommand{\tensor}[1]{\boldsymbol{\mathscr{\MakeUppercase{#1}}}} 

\newcommand{\tE}{\tensor{E}}

\newcommand{\tW}{\tensor{W}}

\newcommand{\tY}{\tensor{Y}}

\title{Cram\'er-Rao-Induced Bounds for
CANDECOMP/PARAFAC tensor decomposition}
\author
{Petr Tichavsk\'{y}$^1$, Anh Huy Phan$^{2}$, and Zbyn\v{e}k
Koldovsk\'{y}$^{1,3}$}

\begin{document}

\maketitle

\footnotetext{This work was supported by the Grant Agency of the
Czech Republic through the project 102/09/1278.
\\$^1$Institute of Information Theory and Automation,
Pod vod\'{a}renskou v\v{e}\v{z}\'{\i} 4, P.O.Box 18,
182 08 Prague 8, Czech Republic. E-mail: tichavsk@utia.cas.cz.\\
$^2$Brain Science Institute, RIKEN, Wakoshi, Japan. E-mail:
phan@brain.riken.jp.\\$^3$Faculty of Mechatronic and
Interdisciplinary Studies, Technical University of Liberec,
Studentsk\'a 2, 461 17 Liberec, Czech Republic. E-mail:
zbynek.koldovsky@tul.cz.}

\begin{abstract}
This paper presents a Cram\'er-Rao lower bound (CRLB) on the
variance of unbiased estimates of factor matrices in Canonical
Polyadic (CP) or CANDECOMP/PARAFAC (CP) decompositions of a tensor
from noisy observations, (i.e., the tensor plus a random Gaussian
i.i.d. tensor).  A novel expression is derived for a bound on the
mean square angular error of factors along a selected dimension of
a tensor of an arbitrary dimension. The expression needs less
operations for computing the bound, $O(NR^6)$, than the best
existing state-of-the art algorithm, $O(N^3R^6)$ operations, where
$N$ and $R$ are the tensor order and the tensor rank. Insightful
expressions are derived for tensors of rank 1 and rank 2 of
arbitrary dimension and for tensors of arbitrary dimension and
rank, where two factor matrices have orthogonal columns.

The results can be used as a gauge of performance of different
approximate CP decomposition algorithms, prediction of their
accuracy, and for checking stability of a given decomposition of a
tensor (condition whether the CRLB is finite or not).
A novel expression is derived
for a Hessian matrix needed in popular damped Gauss-Newton method
for solving the CP decomposition of tensors with missing elements.
Beside computing the CRLB for these tensors the expression may
serve for design of damped Gauss-Newton algorithm for the
decomposition.

\end{abstract}

\centerline{{\small\bf Index Terms}}

{\small Multilinear models; canonical polyadic decomposition;
Cram\'er-Rao lower bound; stability; uniqueness}

\section{Introduction}
\comment{Sometimes you us CP, and sometime you use 'CP decomposition';
I left out CP of the abstract because of this, then you proceed with just
CP but occasionally bring back 'CP decomposition' which may be redundant,
pick what sounds better or is excepted from the literature and stick with it.
This is the problem of 'subtle variation' where author use slightly different
terminology throughout unintentionally, and do not give the readers more insight}

Order-3 and higher-order data arrays need to be analyzed in
diverse research areas such as chemistry, astronomy, and
psychology \cite{Bro1}--\cite{Bro2}. The analyses can be done
through finding multi-linear dependencies among elements within
the arrays. The most popular model is Parallel factor analysis
(PARAFAC), also called Canonical decomposition (CANDECOMP) or CP,
which is an extension of a low rank decomposition of matrices to
higher-way arrays, usually called tensors. In signal processing,
the tensor decompositions have become popular for their usefulness
in blind source separation \cite{nmf}.

Note that a best-fitting CP decomposition may not exist for some
tensors. In that case, trying to find a best-fitting CP
decomposition results in diverging factors
\cite{bestMNE1,bestMNE2}. This paper is focussed on studying CP
decompositions of a noisy observations of tensors, which admit an
exact CP decomposition. The decomposition of the noiseless tensor
is taken as a {\em ground truth} for computing errors.

An important issue is the essential uniqueness of CP decomposition
as it entails identifiability of the model (the factor matrices)
from the tensor. The adjective ``essential" means that the model
is unique up to a scale and permutation ambiguity, which is
inherent to the problem. Initial works in the field can be traced
back in 70's in works of Harshman \cite{Hars1,Hars2}. A popular
sufficient condition for the uniqueness was derived by Kruskal in
\cite{Kruskal}. Recently, the problem has been addressed again,
namely by Stegeman, Ten Berge, De Lathauwer, Jiang, Sidiropoulos
et al.; see \cite{222}-\cite{stegeman12b}.

This paper is focussed on stability of the CP decomposition rather
than on the uniqueness. By stability we mean existence of a finite
Cram\'er-Rao bound in a stochastic set-up, where tensor elements
are corrupted by additive Gaussian-distributed noise. Relation of
this kind of stability to a deterministic stability and to the
uniqueness was studied in \cite{Basu}. It is not true, in general,
that stability of a solution of a nonlinear problem implies
uniqueness of the solution. For example, there might always be a
permutation or sign ambiguity. It is yet an open theoretical
question if stability of the CP tensor decomposition problem
implies its {\em essential} uniqueness. Regardless of the missing
link to identifiability, the stability is an interesting concept
which is worth to be studied, because different kind of noise is
very common.

In general, in order to evaluate performance of a tensor
decomposition, the approximation error between the data tensor and
its approximate is commonly used. Unfortunately, such measure does
not imply quality of the estimated components. In practice, in
some difficult scenarios such as decomposition of tensor with
linear dependency among components of factor matrices, or large
difference in magnitude between components \cite{Tomasi2,paatero},
most CP algorithms explained the data tensor at almost identical
fit, but only few algorithms can accurately retrieve the hidden
components from the tensor \cite{SIAM,Tomasi2}. In order to verify
theoretically the quality of the estimated components and evaluate
robustness of an algorithm, an appropriate measure is an essential
prerequisite. The squared angular error between the estimated
component and its original one is such a measure
\cite{stability,udsep}. Working with angular errors is practical,
because the scaling ambiguity does not play a role. Only the
permutation ambiguity has to be solved in practical examples,
because order of the factor can be quite arbitrary.

Cram\'er-Rao lower bound for CP decomposition was first studied in
\cite{CRB}, and later, a more compact asymptotic expression was
derived in \cite{CRB2} for tensors of order 3 appearing in
wireless communications. A non-asymptotic (exact) CRLB-induced
bound (CRIB) on squared angular deviation of columns of the factor
matrices with respect to their nominal values has been studied in
\cite{stability}. Similar results for symmetric tensors are
derived in \cite{istability}. Nevertheless, the study is limited
to the case of three-way tensors. In the general case, CRIB can
be, indeed, calculated through the approximate Hessian which is
often huge, and is impractical to directly invert. Note that such
task normally costs $O(R^3T^3)$ where $T = \sum_n I_n$. Seeking a
cheaper method for CRIB is a challenge to made it applicable.

This paper presents new CRIB expressions for tensors of arbitrary
dimension and rank, and specialized expressions for rank 1 and
rank 2 tensors. The results rely on compact expressions for
Hessian of the problem derived in \cite{SIAM}. Alternative
expressions for the Hessian exist in \cite{Tendeiro}. Note,
however, that unlike \cite{SIAM}, this paper presents different
expressions for inverse of the Hessian, which have lower
computational complexity. In particular, complexity of inversion
of the Hessian is reduced from $O(N^3R^6)$ operations to
$O(NR^6)$, where $N$ and $R$ are the tensor order and the tensor
rank, respectively. 

On basis of new discovered properties of the CRIB, we established
connection between theoretical and practical results in CPD:
\begin{itemize}
\item Stability of CPD for rank-1 and rank-2 tensors of arbitrary
dimension. \item The work may serve as theoretical support for a
novel CP decomposition algorithm through tensor reshaping
\cite{3w2Nw}, which was designed to decompose high-dimensional and
high-order tensors. In particular, it appears that higher-order
orthogonally constrained CPD \cite{[42],[43],[44],Saad09} can be
decomposed efficiently through tensor unfolding.
\item Stability when factor matrices occur linear dependence
problem and especially the rank-overlap problem
\cite{Bro1,stegeman12b,[43]}. The problem is related to a variant
of CPD for linear dependent loadings which was investigated in
chemometric data and in flow injection analysis \cite{Bro1,[43]}.
A partial uniqueness condition of the related model is discussed
in \cite{stegeman12b}. \item CP decomposition of tensors with
missing entries, which is quite frequent in practice, is
addressed. An approximate Hessian for this case is derived, which
is the core for the damped Gauss-Newton algorithm for the
decomposition. \item A maximum tensor rank, given dimension of the
tensor, which admits a stable decomposition is discussed.
\end{itemize}
The paper is organized as follows. Section II presents the main
result, the Cram\'er-Rao induced bound on angular error of one
factor vector in full generality. In Section III, this result is
specialized for tensors of rank 1 and rank 2, and for the case
when two factor matrices have mutually orthogonal columns. Section
IV is devoted to a possible application of the bound:
investigation of loss of accuracy of the tensor decomposition when
the tensor is reshaped to a lower-dimensional form. Section V
deals with the bound for tensors with missing entries, Section VI
contains examples -- CRIB computed for CP decomposition of a
fluorescence tensor, stability of the tensor investigated by Brie
{\em et al}, and a discussion of a maximum stable rank given the
tensor dimension. Section VII concludes the paper.

\section{Presentation of the CRIB}
\subsection{Cram\'er-Rao bound for CP decomposition}

Let $\tY$ be an $N-$ way tensor of dimension $I_1\times I_2\times
\ldots\times I_N$. The tensor is said to be of rank $R$, if $R$ is
the smallest number of rank-one tensors which admit the
decomposition of $\tY$ of the form
\begin{equation}
  \tY =    \sum\limits_{r = 1}^R {\ba^{(1)}_{r}
  \circ \ba^{(2)}_{r} \circ  \ldots  \circ \ba^{(N)}_{r}}
  \label{rank}
\end{equation}
where $\circ$ denotes the outer vector product, $\ba^{(n)}_r$,
$r=1,\ldots,R$, $n=1,\ldots,N$ are vectors of the length $I_n$
called factors. The tensor in (\ref{rank}) can be characterized by
$N$ factor matrices $\bA_n=[\ba^{(n)}_1,
\ba^{(n)}_2,\ldots,\ba^{(n)}_R]$ of the size $I_n\times R$ for
$n=1,\ldots,N$. Sometimes (\ref{rank}) is referred to as a Kruskal
form of a tensor \cite{Kolda}.

In practice, CP decomposition of a given rank ($R$) is used as an
approximation of a given tensor, which can be a noisy observation
${\hat\tY}$ of the tensor $\tY$ in (\ref{rank}).
Owing to the symmetry of (1), we can focus on estimating the first
factor matrix $\bA_1$, without any loss of generality, and we can
assume that all other factor matrices have
columns of unit norm. 
Then the ``energy" of the parallel factors is determined by the squared
Euclidean norm of columns of $\bA_1$.

It  is common to assume that the noise has a zero mean Gaussian
distribution with variance $\sigma^2$, and is independently added
to each element of the tensor in (\ref{rank}).

Let a vector parameter $\thetab$ containing all parameters of our
model be arranged as
\begin{equation}\label{thetab}
\thetab=[({\tt vec}\,\bA_1)^T,\ldots,({\tt vec}\,\bA_N)^T]^T~.
\end{equation}
The maximum likelihood solution for $\thetab$ consists in
minimizing the least squares criterion
\begin{equation}
{\cal Q}(\thetab)=\|{\hat\tY}-\tY(\thetab)\|_F^2
\end{equation}
where $\|\cdot\|_F$ stands for the Frobenius norm.

We wish to compute the Cram\'er-Rao lower bound for estimating
$\thetab$. In general, for this estimation problem, the CRLB is
given as the inverse of the Fisher information matrix, which is
equal to \cite{stability}
\begin{equation}
{\bf F}(\thetab)=\frac{1}{\sigma^2}\,\bJ^T(\thetab)\bJ(\thetab)
\end{equation}
where $\bJ(\thetab)$ is the Jacobi matrix (matrix of the
first-order derivatives) of ${\cal Q}(\thetab)$ with respect to
$\thetab$. In other words, the Fisher information matrix is
proportional to the approximate Hessian matrix of the criterion, ${\bf
H}(\thetab)=\bJ^T(\thetab)\bJ(\thetab)$.

Let $\boldsymbol \Gamma_{nm}$ denote the Hadamard (elementwise)
product of matrices $\bC_k=\bA_k^T\bA_k, k\in
\{1,\ldots,N\}-\{n,m\}$,
\begin{eqnarray}
{\bGamma}_{nm} &=&   \bigcircledast\limits_{k\neq n,m}{\bC_k} \,
,\qquad \bC_k = \bA_k^T \bA_k~.\label{defC}
\end{eqnarray}

\noindent{\bf Theorem 1} \cite{SIAM}: The Hessian $\bH$ can be
decomposed into low rank matrices under the form as
\begin{equation}
    \bH = \bG + \bZ \, \bK \, \bZ^T\label{hessi}
\end{equation}
where $\bK =  \left[ \bK_{nm}\right]_{n,m=1}^N$ contains
submatrices $\bK_{nm}$ given by
\begin{equation}
    \bK_{nm} = (1-\delta_{nm})
            \bP_{R} \, \dvec\,(\bGamma_{nm}) \label{defK}
\end{equation}
$\bP_{R}$ is the permutation matrix of dimension $R^2 \times R^2$
defined in \cite{SIAM} such that ${\tt{vec}}\,{\bM} = \bP_{R} \,
{\tt{vec}}\,({\bM^T})$ for any $R\times R$ matrix $\bM$, and
$\delta_{nm}$ is the Kronecker delta, and ${\tt dvec}(\bM)$ is a
short-hand notation for ${\tt diag}({\tt vec}(\bM))$, i.e. a
diagonal matrix containing all elements of a matrix $\bM$ on its
main diagonal.  Next,
\begin{eqnarray}
    \bG &=&  {\tt{bdiag}}\left(  {\boldsymbol\Gamma}_{nn} \otimes
    \bI_{I_n}
            \right)_{n=1}^{N}
\end{eqnarray}
and
\begin{equation}
\bZ = {\tt{bdiag}}\left(\bI_R \otimes
\bA_n\right)_{n=1}^{N}\label{defZ}
\end{equation}
where $\otimes$ denotes the Kronecker product, $\bI_{I_n}$ is an
identity matrix of the size $I_n\times I_n$, and
${\tt{bdiag}}(\cdot)$ is a block diagonal matrix with the given
blocks on its diagonal. Note that the Hessian $\bH$ in
(\ref{hessi}) is rank deficient because of the scale ambiguity of
columns of factor matrices \cite{paatero,Tomasi}. It has dimension
$\left(R\sum_n I_n\right) \times \left(R\sum_n I_n\right)$ but its
rank is at most $R\sum_n I_n-(N-1)R$.

A regular (reduced) Hessian can be obtained from $\bH$ by deleting
$(N-1)R$ rows and corresponding columns in $\bH$, because the
estimation of one element in the vectors $\ba_r^{(n)}$,
$r=1,\ldots,R$, $n=2,\ldots,N$ can be skipped. The reduced Hessian
may have the form
\begin{equation}
\bH_E = \bE\bH\bE^T\label{EHE}
\end{equation}
where
\begin{equation}
\bE = {\tt{bdiag}}\left(\bI_{RI_1}, \bI_R\otimes\bE_2,\ldots,
\bI_R\otimes\bE_N\right)\label{defE}
\end{equation}
and $\bE_n$ is an $(I_n-1)\times I_n$ matrix of rank $I_n-1$. For
example, one can put $\bE_n=[{\bf 0}_{(I_n-1)\times
1}\,\bI_{I_n-1}]$ for $n=2,\ldots,N$. With this definition of
$\bE_n$, $\bH_E$ is a Hessian for estimating the first factor
matrix $\bA_1$ and all other vectors $\ba_r^{(n)}$,
$r=1,\ldots,R$, $n=2,\ldots,N$ without their first elements. In
the sequel, however, we use a different definition of $\bE_n$.
Note that each $\bE_n$ can be quite arbitrary, together facilitate
a regular transformation of nuisance parameters, which does not
influence CRLB of the parameter of interest.

The CRLB for the first column of $\bA_1$, denoted simply as
$\ba_1$, is defined as $\sigma^2$ times the left-upper submatrix
of $\bH_E^{-1}$ of the size $I_1\times I_1$,
\begin{equation}
\mbox{CRLB}(\ba_1)=\sigma^2\,[\bH_E^{-1}]_{1:I_1,1:I_1}~.\label{crlb}
\end{equation}
Substituting (\ref{hessi}) in (\ref{EHE}) gives
\begin{equation}
\bH_E=\bG_E+\bZ_E\bK\bZ_E^T
\end{equation}
where $\bG_E=\bE\bG\bE^T$ and $\bZ_E=\bE\bZ$. Inverse of $\bH_E$
can be written using a Woodbury matrix identity \cite{MIL} as
\begin{equation}
\bH_E^{-1}=\bG_E^{-1}-\bG_E^{-1}\bZ_E\bK(\I_{NR^2}+\bZ_E^T\bG_E^{-1}\bZ_E\bK)^{-1}\bZ_E^T\bG_E^{-1}
\end{equation}
provided that the involved inverses exist.

Next,
\begin{eqnarray}
\bG_E &=& {\tt{bdiag}}\left(  {\boldsymbol\Gamma}_{11} \otimes
\bI_1,{\boldsymbol\Gamma}_{22} \otimes
(\bE_2\bE_2^T),\ldots,{\boldsymbol\Gamma}_{NN} \otimes
(\bE_N\bE_N^T)\right)\\
\bG_E^{-1} &=& {\tt{bdiag}}\left(({\boldsymbol\Gamma}_{11})^{-1}
\otimes \bI_1,\bGamma_{22}^{-1} \otimes
(\bE_2\bE_2^T)^{-1},\ldots,{\boldsymbol\Gamma}_{NN}^{-1} \otimes
(\bE_N\bE_N^T)^{-1}\right)~.\label{defGin}
\end{eqnarray}
Put
\begin{eqnarray}
\bPsi &=& \bZ_E^T\bG_E^{-1}\bZ_E\label{defPsi0}\\
\bB &=& \bK(\I_{NR^2}+\bPsi\bK)^{-1} \label{defB}
\end{eqnarray}
and let $\bB_0$ be the upper--left $R^2\times R^2$ submatrix of $\bB$, symbolically $\bB_0=\bB_{1:R^2,1:R^2}$.
Finally, let $g_{11}$ and $\bg_{1,:}$ be the upper--left
element and the first row of ${\boldsymbol\Gamma}_{11}^{-1}$,
respectively. Then
\begin{eqnarray}
[\bH_E^{-1}]_{1:I_1,1:I_1} &=& [\bG_E^{-1}]_{1:I_1,1:I_1} +
[\bG_E^{-1}\bZ_E]_{1:I_1,1:R^2}\bB_0[\bG_E^{-1}\bZ_E]_{1:I_1,1:R^2}^T\nonumber\\
&=& g_{11}\bI_{I_1} + \left(\bg_{1,:}
\otimes\bA_1\right)\bB_0\left(\bg_{1,:}
\otimes\bA_1\right)^T~.\label{dE}
\end{eqnarray}
The CRLB represents a lower bound on the error covariance matrix
$\mbox{E}[(\hat{\ba}_1-\ba_1)(\hat{\ba}_1-\ba_1)^T]$ for any
unbiased estimator of $\ba_1$. The bound is asymptotically tight
in the case of Gaussian noise and least squares estimator, which
is equivalent to maximum likelihood estimator, under the
assumptions that the permutation ambiguity has been solved out
(order of the estimated factors was selected to match the original
factors) and scaling of the estimator is in accord with the
selection of the matrix $\bE$.

\subsection{Cram\'er-Rao-induced bound for angular error}

CRLB($\ba_1$) considered in the previous subsection is a matrix.
In applications it is practical to characterize the error of the
factor $\ba_1$ in the decomposition by a scalar quantity. In
\cite{udsep} it was proposed to characterize the error by an angle
between the true and the estimated vector, and compute a
Cram\'er-Rao-induced bound (CRIB) for the squared angle. The CRIB
may serve a gauge of achievable accuracy of estimation/CP
decomposition. Again, it is an asymptotically (in the sense of
variance of the noise going to zero) tight bound on the angular
error between an estimated and true factor.

The angle $\alpha_1$ between the true factor $\ba_1$ and its
estimate $\hat{\ba}_1$ obtained through the CP decomposition is
defined through its cosine
\begin{equation}
\cos\alpha_1 = \frac{\ba_1^T\hat{\ba}_1}{\|\ba_1\|\,\|\hat{\ba}_1\|}~.
\end{equation}
The Cram\'er-Rao induced bound for the squared angular error
$\alpha_1^2$ [radians$^2$] will be denoted CRIB($\ba_1$) in the
sequel. CRIB($\ba_1$) in decibels (dB) is then defined as
$-10\log_{10}[\mbox{CRIB}(\ba_1)]\,[\mbox{dB}]$.

Before computing CRIB($\ba_1$) we present another interpretation of this quantity.
Let the estimate $\hat{\ba}_1$ be decomposed into a sum of a scalar multiple of $\ba_1$ and a reminder, which is orthogonal to $\ba_1$,
\begin{equation}
\hat{\ba}_1=\beta\ba_1+\br_1
\end{equation}
where $\beta=\ba_1^T\hat{\ba}_1/\|\ba_1\|^2$ and $\br_1=\hat{\ba}_1-\beta\ba_1$.
Then, the Distortion-to-Signal Ratio (DSR) of the estimate $\hat{\ba}_1$ can be defined as
\begin{equation}
\mbox{DSR}(\hat{\ba}_1)=\frac{\|\br_1\|^2}{\beta^2\|\ba_1\|^2}~.
\end{equation}
A straightforward computation gives
\begin{equation}
\mbox{DSR}(\hat{\ba}_1)=\frac{1-\cos^2\alpha_1}{\cos^2\alpha_1}\approx
\alpha_1^2~.\label{DSR}
\end{equation}
The approximation in (\ref{DSR}) is valid for small $\alpha_1^2$.
We can see that CRIB($\ba_1$) serves not only as a bound on the
mean squared angular estimation error, but also as a bound on the
achievable Distortion-to-Signal Ratio. \vspace{5mm}

\noindent{\bf Theorem 2} \cite{udsep}: Let CRLB($\ba_1$) be the
Cram\'er-Rao bound on covariance matrix of unbiased estimators of
$\ba_1$. Then the Cram\'er--Rao--induced bound on the squared
angular error between the true and estimated vector is
\begin{equation}
\mbox{CRIB}(\ba_1)= \frac{\mbox{tr}[\Pi^\bot_{{\bf
a}_1}\mbox{CRLB}({\bf a}_1)]}{\|{\bf a}_1\|^2}\label{crib}
\end{equation}
where
\begin{eqnarray}
\Pi^\bot_{{\bf a}_1}=\I_{I_1}-{\bf a}_1{\bf a}_1^T/\|{\bf a}_1\|^2
\end{eqnarray}
is the projection operator to the orthogonal complement of ${\bf
a}_1$ and tr[.] denotes trace of a matrix.

\noindent{\bf Proof}: A sketch of a proof can be found in
\cite{udsep}. It is based on analysis of a mean square angular
error of a maximum likelihood estimator, which is known to be
asymptotically tight (achieving the Cram\'er-Rao bound). Note that
a conceptually more straightforward but longer proof would be
obtained through the formula for CRLB on a transformed parameter,
see e.g., Theorem 3.4 in \cite{Porat}. In particular,
\begin{equation}
\mbox{CRIB}(\ba_1)= \bG_a(\ba_1)\mbox{CRLB}({\bf
a}_1)\bG_a^T(\ba_1) \label{crib2}
\end{equation}
where $\bG_a(\hat{\ba}_1)$ is the Jacobi matrix of the mapping
representing the angular error as a function of the estimate
$\hat{\ba}_1$.

\noindent{\bf Theorem 3}: The CRIB($\ba_1$) can be written in the
form
\begin{eqnarray}
\mbox{CRIB}(\ba_1)= \frac{\sigma^2}{\|{\bf a}_1\|^2}\left\{
(I_1-1)g_{11}-\mbox{tr}\left[\bB_0\left((\bg_{1,:}^T\bg_{1,:})
\otimes \bX_1 \right)\right]\right\} \label{result}
\end{eqnarray}
where $\bB_0$ is the submatrix of $\bB$ in (\ref{defB}),
$\bB_0=\bB_{1:R^2,1:R^2}$,
\begin{equation}
\bX_n=\bC_n-\frac{1}{\bC_{11}^{(n)}}\bC_{:,1}^{(n)}\bC_{:,1}^{(n)\,T}\label{defX}
\end{equation}
for $n=1,\ldots,N$, $\bC_{11}^{(n)}$ and $\bC_{:,1}^{(n)}$ denote
the upper--right element and the first column of $\bC_n$,
respectively, and $\bPsi$ in the definition of $\bB$ takes, for a
special choice of matrices $\bE_n$, the form
\begin{equation}
\bPsi={\tt{bdiag}}\left( \bGamma_{11}^{-1} \otimes
\bC_1,\bGamma_{22}^{-1} \otimes \bX_2,\ldots,\bGamma_{NN}^{-1}
\otimes \bX_N\right)~.\label{defPsi2}
\end{equation}
 \noindent{\bf Proof}: Substituting (\ref{crlb}) and (\ref{dE})
into (\ref{crib}) gives, after some simplifications,
\begin{eqnarray}
\mbox{CRIB}(\ba_1)&=& \frac{\sigma^2}{\|{\bf
a}_1\|^2}\mbox{tr}\left[\Pi^\bot_{{\bf a}_1}\left(g_{11}\bI_{I_1}
- \left(\bg_{1,:} \otimes\bA_1\right)\bB_0\left(\bg_{1,:}
\otimes\bA_1\right)^T\right)\right]\nonumber\\&=&
\frac{\sigma^2}{\|{\bf
a}_1\|^2}\left\{(I_1-1)g_{11}-\mbox{tr}\left[\Pi^\bot_{{\bf
a}_1}\left(\bg_{1,:} \otimes\bA_1\right)\bB_0\left(\bg_{1,:}
\otimes\bA_1\right)^T\right]\right\}\nonumber\\&=&
\frac{\sigma^2}{\|{\bf
a}_1\|^2}\left\{(I_1-1)g_{11}-\mbox{tr}\left[\bB_0\left((\bg_{1,:}^T\bg_{1,:})
\otimes \left(\bA_1^T\Pi^\bot_{{\bf
a}_1}\bA_1\right)\right)\right]\right\}~.
\label{crib3}
\end{eqnarray}
This is (\ref{result}), because
\begin{equation}
\bA_1^T\Pi^\bot_{{\bf a}_1}\bA_1
=\bC_1-\frac{1}{\bC_{11}^{(1)}}\bC_{:,1}^{(1)}\bC_{:,1}^{(1)\,T}=\bX_1~.
\end{equation}
%
Next, assume that $\bE$ is defined as in (\ref{defE}), but $\bE_n$
are arbitrary full rank matrices of the dimension $(I_n-1)\times
I_n$. Then, combining (\ref{defPsi0}), (\ref{defZ}), (\ref{defE})
and (\ref{defGin}) gives
\begin{equation}
\bPsi=\bZ_E^T\bG_E^{-1}\bZ_E={\tt{bdiag}}\left( \bGamma_{11}^{-1}
\otimes \bC_1,\bGamma_{22}^{-1} \otimes
\widetilde{\bX}_2,\ldots,\bGamma_{NN}^{-1} \otimes
\widetilde{\bX}_N\right)\label{defPsi}
\end{equation}
where
\begin{equation}
\widetilde{\bX}_n=\bA_n^T\bE_n^T(\bE_n\bE_n^T)^{-1}\bE_n\bA_n
\end{equation}
for $n=2,\ldots,N$. Note that the expression
$\bE_n^T(\bE_n\bE_n^T)^{-1}\bE_n$ is an orthogonal projection
operator to the columnspace of $\bE_n^T$. If $\bE_n$ is chosen as
the first $(I_n-1)$ rows of
\begin{eqnarray}
\Pi^\bot_{{\bf a}_1^{(n)}}=\I_{I_n}-{\bf a}_1^{(n)}{\bf
a}_1^{(n)\,^T}/\|{\bf a}_1^{(n)}\|^2
\end{eqnarray}
then $\bE_n^T(\bE_n\bE_n^T)^{-1}\bE_n=\Pi^\bot_{{\bf a}_1^{(n)}}$
and consequently $\widetilde{\bX}_n=\bA_n^T\Pi^\bot_{{\bf
a}_1^{(n)}}\bA_n=\bX_n$. \hfill \rule{2mm}{2mm}

Note that the first row and the first column of $\bX_n$ are zero.

\noindent{\bf Theorem 4}: Assume that all elements of the matrices
$\bC_n$ in (\ref{defC}) are nonzero. Then, the matrix $\bB_0$ in
Theorem 3 can be written in the form
\begin{equation}
\bB_0=[-\bI_{R^2}+\bV(\bI_{R^2}+\bV)^{-1}]\bY\label{result4}
\end{equation}
where
\begin{eqnarray}
\bV&=&\bW-\bY (\bGamma_{11}^{-1}\otimes\bC_1)\\
\bW&=& \bP_R \sum_{n=2}^N {\tt dvec}(\bGamma_{1n})
\bS_n^{-1}(\bGamma_{nn}^{-1}\otimes\bX_n)
{\tt dvec}(\bC_1\oslash\bC_n)\label{defW}\\
\bY&=&\bP_R \sum_{n=2}^N {\tt dvec}(\bGamma_{1n})
\bS_n^{-1}(\bGamma_{nn}^{-1}\otimes\bX_n) \bP_R{\tt dvec}(\bGamma_{1n})\\
\bS_n&=& \bI_{R^2}
-(\bGamma_{nn}^{-1}\otimes\bX_n){\tt
dvec}(\bGamma_{nn}\oslash\bC_n)\bP_R,\qquad
n=2,\ldots,N~.\label{defSn}
\end{eqnarray}
In (\ref{defW}) and (\ref{defSn}), ``$\oslash$" stands for the element-wise division. \\
\noindent{\bf Proof:} See Appendix B.\\

Note that in place of inverting the matrix $\bB$ of the size
$NR^2\times NR^2$, Theorem 4 reduces the complexity of the CRIB
computation to $N$ inversions of the matrices of the size
$R^2\times R^2$. The Theorem can be extended to computing the
inverse of the whole Hessian in $O(NR^6)$ operations, see
\cite{icassp2013}.


Finally, note that the assumption that elements of $\bC_n$ must not
be zero is not too restrictive. Basically, it means that no pair
of columns in the factor matrices must be orthogonal. The
Cram\'er-Rao bound does not exhibit any singularity in these
cases, and is continuous function of elements of $\bC_n$. If some
element of $\bC_n$ is closer to zero than say $10^{-5}$, it is
possible to increase its distance from zero to that value, and the
resultant CRIB will differ from the true one only slightly.

 \noindent{\bf Theorem 5} (Properties of the CRIB)
\begin{enumerate}
\item The CRIB in Theorems 3 and 4 depends on the factor matrices
$\bA_n$ only through the products $\bC_n=\bA_n^T\bA_n$. \item The
CRIB is inversely proportional to the signal-to-noise ratio (SNR)
of the factor of the interest (i.e. $\|\ba_1\|^2/(\sigma^2 I_1)$) and
independent of the SNR of the other factors,
$\|\ba_r\|^2/(\sigma^2 I_r)$, $r=2,\ldots,R$.
\end{enumerate}
\noindent{\bf Proof:} Property 1 follows directly from Theorem 3.
Property 2 is proven in Appendix C.

\section{Special cases}


\subsection{Rank 1 tensors}

In this case, the matrix $\bX_1$ is zero, and
\begin{eqnarray}
\mbox{CRIB}(\ba_1)= \frac{\sigma^2}{\|\ba_1\|^2}\,(I_1-1)g_{11}
=\frac{\sigma^2}{\|\ba_1\|^2}\,(I_1-1)~. \label{rank1}
\end{eqnarray}
In (\ref{rank1}), $g_{11}=1$ due to the convention that the factor
matrices $\bA_n$, $n\ge 2$, have columns of unit norm. The result
(\ref{rank1}) is in accord with Harshman's early results on
uniqueness of rank-1 tensor decomposition \cite{Hars2}.

\subsection{Rank 2 tensors}

Consider the scaling convention that all factor vectors except the
first factor have unit norm. Let $c_n$, $|c_n|\leq 1$, be 
defined as
\begin{eqnarray}
c_n=\left\{\begin{array}{lcl}(\ba_1^{(n)})^T\ba_2^{(n)} &
\mbox{for} &
n=2,\ldots,N\\
(\ba_1^{(1)})^T\ba_2^{(1)}/(\|\ba_1^{(1)}\|\,\|\ba_2^{(1)}\|) &
\mbox{for} & n=1~.\end{array}\right.
\end{eqnarray}
It follows from Theorem 5 that the CRIB on $\ba_1$ is a function
of $c_1,\ldots,c_N$ multiplied by $\sigma^2/\|\ba_1\|^2$. It is
symmetric function in $c_2,\ldots,c_N$ and possibly nonsymmetric
in $c_1$. A closed form expression for the CRIB in the special
case is subject of the following theorem.

\noindent{\bf Theorem 6} It holds for rank 2 tensors
\begin{eqnarray}
\mbox{CRIB}(\ba_1)=
\frac{\sigma^2}{\|\ba_1\|^2}\,\frac{1}{1-h_1^2}\left[I_1-1 +
\frac{(1-c_1^2)h_1^2[y^2+z-h_1^2z(z+1)]}{(1 - c_1y -
h_1^2(z+1))^2-h_1^2(y + c_1z)^2}\right]\label{result31}
\end{eqnarray}
where
\begin{eqnarray}
h_n &=& \prod_{2\leq k\ne n}^N c_n \qquad\mbox{for}\quad n=1,\ldots,N\\
y &=& - c_1\,\sum_{n=2}^N \frac{h_n^2(1-c_n^2)}{c_n^2-h_n^2c_1^2}
\label{exp1}\\ z &=& \,\sum_{n=2}^N
\frac{1-c_n^2}{c_n^2-h_n^2c_1^2}~.\label{exp3}
\end{eqnarray}
\noindent{\bf Proof:} See Appendix D.

Note that the expressions (\ref{exp1})-(\ref{exp3}) contain, in their
denominators, terms $c_n-h_nc_1$. If any of these terms goes to
zero, then quantities $y$ and $z$ go to infinity. In despite of this,
the whole CRIB remain finite, because $y$ and $z$ appear both in
the numerator and denominator in (\ref{result31}).

For example, for order-3 tensors ($N=3$) we get (using e.g.,
Symbolic Matlab or Mathematica)
\begin{eqnarray}
\mbox{CRIB}_{N=3}(\ba_1)=
\frac{\sigma^2}{\|\ba_1\|^2}\,\frac{1}{1-h_1^2}\left[I_1-1+
\frac{c_2^2}{1-c_2^2}+\frac{c_3^2}{1-c_3^2}\right]~.
\end{eqnarray}
The above result coincides with the one derived in
\cite{stability}. As far as the stability is concerned, the CRIB
is finite unless either the second or third factor have co-linear
columns.
Note that the fact that the CRIB for $\ba_1$ does not
depend on $c_1$ can be linked to the uni-mode uniqueness
conditions presented in \cite{stegeman12b}.

For $N=4$, the similar result is hardly tractable. Unlike the case
$N=3$, the result depends on
$c_1$. A closer inspection of the
result shows that the CRIB, as a function of $c_1$, achieves its
maximum at $c_1=0$, and minimum at $c_1=\pm 1$. Therefore we shall
treat these two limit cases separately.

We get
\begin{eqnarray}
\mbox{CRIB}_{N=4,c_1=0}(\ba_1)&=&
\frac{\sigma^2}{\|\ba_1\|^2}\,\frac{1}{1-h_1^2}\left[
I_1-1+\frac{c_2^2c_3^2+c_2^2c_4^2+c_3^2c_4^2-3c_2^2c_3^2c_4^2}
{2c_2^2c_3^2c_4^2 - c_2^2c_3^2 - c_2^2c_4^2 -
c_3^2c_4^2 + 1}\right]\\
\mbox{CRIB}_{N=4,c_1=\pm 1}(\ba_1)&=&\left\{\begin{array}{ll}
\frac{\sigma^2}{\|\ba_1\|^2}\,\frac{I_1-1}{1-h_1^2}&\mbox{for}\quad
(|c_2|<1) \&(|c_3|<1) \&
(|c_4|<1)\\
\frac{\sigma^2}{\|\ba_1\|^2}\,\frac{1}{1-h_1^2}
\left[I_1-1+\frac{c_2^2+c_3^2-2c_2^2c_3^2}{(1-c_2^2)(1-c_3^2)}\right]
&\mbox{for} \quad|c_4|=1\\
\\
\frac{\sigma^2}{\|\ba_1\|^2}\,\frac{1}{1-h_1^2}\left[I_1-1+\frac{c_2^2+c_4^2-2c_2^2c_4^2}{(1-c_2^2)
(1-c_4^2)}\right]
&\mbox{for}\quad |c_3|=1\\
\\
\frac{\sigma^2}{\|\ba_1\|^2}\,\frac{1}{1-h_1^2}\left[I_1-1+
\frac{c_3^2+c_4^2-2c_3^2c_4^2}{(1-c_3^2)(1-c_4^2)}\right]
&\mbox{for}\quad |c_2|=1~.\end{array}\right.\label{3cases}
\end{eqnarray}
As far as the stability is concerned, we can see that the CRIB is
always finite unless two of the factor matrices have co-linear
columns.

Similarly, for a general $N$, we have for $c_1=0$
\begin{eqnarray}
\mbox{CRIB}_{c_1=0}(\ba_1)&=&\frac{\sigma^2}{\|\ba_1\|^2}\,
\frac{1}{1-h_1^2}\left[I_1-1+\frac{h_1^2z} {1-h_1^2(z+1)}\right]~.
\end{eqnarray}

\subsection{A case with two factor matrices having orthogonal columns}

This subsection presents a closed-form CRIB for a tensor of a
general order and rank, provided that two of its factor matrices
have mutually orthogonal columns. 
The result cannot be derived from Theorem
5, because assumptions of the theorem are not fulfilled.

\noindent{\bf Theorem 7} When the factor matrices $\bA_1$ and
$\bA_2$ both have mutually orthogonal columns, it holds
\begin{eqnarray}
\mbox{CRIB}(\ba_1)&=&\frac{\sigma^2}{\|\ba_1\|^2}
\left[I_1-1+\sum_{r=2}^R \frac{\gamma_r^2}{1-\gamma_r^2}\right]
\end{eqnarray}
where $\gamma_r=\prod_{n=3}^N(\ba_1^{(n)})^T\ba_r^{(n)}$
for $r=2,\ldots,R$.\\
\noindent{\bf Proof:} See Appendix E.

Theorem 7 represents an important example when a tensor reshaping
(see Section V.A. and \cite{3w2Nw} for more details) enables very
efficient (fast) CP decomposition without compromising accuracy.
It has close connection with orthogonally constrained CPD
\cite{[43],[44],Saad09}.

\section{CRIB for tensors with missing observations}

It happens in some applications, that tensors to be decomposed via
CP have missing entries (some observations are simply missing). In
this case, it is possible to treat stability of the decomposition
through the CRIB as well. The only problem is that it is not
possible to use expressions in Theorems 3-8 in such cases.

Assume that the tensor to be studied is given by its factor
matrices $\bA_1,\ldots,\bA_N$ and a 0-1 ``indicator" tensor $\tW$
of the same dimension as $\tY$, which determines which tensor
elements are available (observed). The task is
to compute CRIB for columns of the factor matrices, like in the
previous sections. The CRIB is computed through the Hessian matrix
$\bH$ as in (12) and (20), but its fast inversion is no longer
possible. The Hessian itself can be computed as in its earlier
definition
\begin{eqnarray}
\bH=\bJ_W^T(\thetab)\bJ_W(\thetab),\qquad
\bJ_W(\thetab)=\frac{\partial{\tt{vec}}(\tY\circledast\tW)}{\partial\thetab}
\label{51}
\end{eqnarray}
where $\thetab$ is the parameter of the model (\ref{thetab}). More
specific expressions for the Hessian can be derived in a
straightforward manner. 

\noindent{\bf Theorem 8}:\label{theo_Hessian_missing} Consider the
Hessian for tensor with missing data as an $N \times N$
partitioned matrix $\bH =  [\bH^{(n,m)}]_{n=1,m=1}^{N,N}$ where
$\bH^{(n,m)} = [\bH^{(n,m)}_{r,s}]_{r=1,s=1}^{R,R} \in {\mathbb
R}^{R I_n \times R I_m}
 $. Then
\begin{eqnarray}
     \bH^{(n,m)}_{r,s} = \begin{cases}
     {\tt{diag}}\left( \tW \, {\bar\times}_{-n}  \left\{\ba^{(1)}_r \circledast \ba^{(1)}_{s}, \cdots, \ba^{(N)}_r \circledast \ba^{(N)}_{s}\right\}\right)\quad, & n =m ,  \\
     (\ba^{(n)}_r  \, \ba^{(m)T}_s) \circledast \left(\tW \, {\bar\times}_{-\{n,m\}}  \left\{\ba^{(1)}_r \circledast \ba^{(1)}_{s}, \cdots, \ba^{(N)}_r \circledast \ba^{(N)}_{s} \right\}\right), \quad & n \ne m 
     \end{cases}
\end{eqnarray}
$\tY {\bar\times}_n {\bf u}_n$ denotes the mode-$n$ tensor-vector product between $\tY$ and ${\bf u}_n$ \cite{nmf},
and
\begin{eqnarray}
\tY {\bar\times}_{-n}  \{{\bf u}\}= \tY \, {\bar\times}_{1}  {{\bf u}}_{1} \cdots {\bar\times}_{n-1}  {{\bf u}}_{n-1}  {\bar\times}_{n+1}  {{\bf u}}_{n+1}  \cdots {\bar\times}_{N}  {{\bf u}}_{N}.
\end{eqnarray}
\noindent{\bf Proof:} See Appendix F.

Theorem 8 can be used either to compute the CRIB for tensors with
missing elements, or for implementing damped Gauss-Newton method
for finding the decomposition in difficult cases, where ALS
converges poorly.

\section{Application and Examples}
\subsection{Tensor decomposition through reshape}

Assume that the tensor to-be decomposed is of dimension $N\geq 4$.
The tensor can be reshaped to a lower dimensional tensor, which is
computationally easier to decompose, so that the first factor
matrix remains unchanged. The topic will be better elaborated in
our next paper \cite{3w2Nw}, in this paper we present only the
main idea on two examples, to demonstrate usefulness of the CRIB.

In the first example, consider $N=4$. The tensor in (\ref{rank})
can be reshaped to an order-3 tensor
\begin{equation}
  \tY_{res} = \sum\limits_{r = 1}^R \ba^{(1)}_{r}  \circ \ba^{(2)}_{r} \circ
   (\ba^{(4)}_{r}\otimes \ba^{(3)}_{r})~.
  \label{rankb}
\end{equation}
Both the original and the re-shaped tensors have the same number
of elements ($I_1I_2I_3I_4$) and the same noise added to them.

The question is, what is the accuracy of the factor matrix of the
reshaped tensor compared to the original one. The latter accuracy
should be worse, because a decomposition of the reshaped tensor
ignores structure of the third factor matrix. The question is, by how
much worse. If the difference were negligible, then it is advised to
decompose the simpler tensor (of lower dimension).

If the tensor has rank one, accuracy of both decompositions is the
same. It is obvious from (29).

Let us examine tensors of rank 2. If the original tensor has
correlations between columns of the factor matrices $c_1$, $c_2$,
$c_3$ and $c_4$, the reshaped tensor has correlations $c_1$,
$c_2$, and $c_3c_4$, respectively. $\mbox{CRIB}(\ba_1)$ of the
reshaped tensor is independent of $c_1$, while CRIB of the
original tensor is dependent on $c_1$, so there is a difference,
in general. The difference will be smallest for $c_1=0$
(orthogonal factors) and largest for $c_1$ close to $\pm 1$
(nearly or completely co-linear factors along the first
dimension).

The smallest difference between $\mbox{CRIB}(\ba_1)$ for the
reshaped tensor and for the original one is
$$
\frac{\sigma^2}{\|\ba_1\|^2}\left[\frac{c_2^2+c_3^2c_4^2-2c_2^2c_3^2c_4^2}{(1-c_2^2)(1-c_3^2c_4^2)}
-\frac{c_2^2c_3^2+c_2^2c_4^2+c_3^2c_4^2-3c_2^2c_3^2c_4^2}
{(1-c_2^2c_3^2c_4^2)(2c_2^2c_3^2c_4^2 - c_2^2c_3^2 - c_2^2c_4^2 -
c_3^2c_4^2 + 1)}\right]
$$
and the largest difference is
$$
\frac{\sigma^2}{\|\ba_1\|^2}\left[\frac{c_2^2+c_3^2c_4^2-2c_2^2c_3^2c_4^2}{(1-c_2^2)(1-c_3^2c_4^2)}\right]=
\frac{\sigma^2}{\|\ba_1\|^2}\left[\frac{c_2^2}{1-c_2^2}+\frac{c_3^2c_4^2}{1-c_3^2c_4^2}\right]~.
$$
We can see that the difference may be large if the second or third
factor matrix of the reshaped tensor has nearly co-linear columns
($c_2^2\approx 1$ or $c_3^2c_4^2\approx 1$)~. For example, for a
tensor with $I_1=5$, $c_1=0,\,c_2=0.99,\,c_3=c_4=0.1$ the loss of
accuracy in decomposing reshaped tensor in place of the original
one is 11.22 dB. If $c_1$ is changed to 1, the loss is only
slightly higher, 11.23 dB. If $c_1=c_2=0$, the loss is 0~dB for
any $c_3,\, c_4$ (compare Theorem 7). If $c_1=1,\,c_2=0$ and
$c_3=c_4=0.99$, the loss is 8.5~dB.

Another example is a tensor of an arbitrary order and rank
considered in Theorem 7. Let this tensor be reshaped to the
order-3 tensor of the size $I_1\times I_2\times (I_3\ldots I_N)$.
Comparing the CRIB$(\ba_1)$ of the original tensor and of the
reshaped tensor shows that these two coincide. It follows that the
decomposition based on reshaping is lossless in terms of accuracy.


\subsection{Amino Acids Tensor}
A data set consisting of five simple laboratory-made samples of
fluorescence excitation-emission (5 samples $\times$ 201 emission
wavelengths $\times$ 61 excitation wavelengths) is considered.
Each sample contains different amounts of tryptophan, tyrosine,
and phenylalanine dissolved in phosphate buffered water. The
samples were measured by fluorescence on a spectrofluorometer
\cite{bro_book}. Hence,  a CP model with $R = 3$ is appropriate to
the fluorescence data.

The tensor was factorized for several possible ranks $R$ using the
fLM algorithm \cite{SIAM}. CRIBs on the extracted components were
then computed with the noise levels deduced from the error tensor
${\tE} = {\tY} - {\hat\tY}$
\begin{equation}
\sigma^2 = \frac{\|{\tY}  - {\hat\tY}\|_F^2}{\prod_n
I_n}.\label{estsig}
\end{equation}
The resultant CRIB's are computed for all columns of all factor
matrices and are summarized in Table 1.

\begin{table}[t]
\centering \caption{Estimated CRIBs [{\tt dB}] on best fit CP
components of fluorescence tensor computed for assumed rank $R =1,
2, 3, 4$}
\begin{tabular*}{1\textwidth}{@{\extracolsep{\fill}}c@{\hspace{1ex}}c@{\hspace{1ex}}c@{\hspace{.5ex}}c@{\hspace{1ex}}c@{\hspace{.5ex}}c@{\hspace{.5ex}}c@{\hspace{1ex}}c@{\hspace{.5ex}}c@{\hspace{.5ex}}c@{\hspace{.5ex}}c@{}}
\hline\hline {\footnotesize{Factor}} & \multicolumn{1}{c}{$R = 1$}
& \multicolumn{2}{c}{$R = 2$} & \multicolumn{3}{c}{$R = 3$} &
\multicolumn{4}{c}{$R = 4$}\\
\cline{2-2}\cline{3-4}\cline{5-7}\cline{8-11} $n$& 1 &  1 & 2 & 1
& 2 & 3& 1 & 2 & 3 & 4\\
\cline{1-1}\cline{2-2}\cline{3-4}\cline{5-7}\cline{8-11}
1&    44.43&   44.44&   41.87&   \bf 64.76&   \bf 61.34&   \bf 64.98&   65.78&   60.96&   65.77&   38.17    \\
2&    {27.44}&   30.28&   {27.71}&  \bf  53.15& \bf   50.17&   \bf 49.60&   54.33&   51.39&   50.87&   {23.29} \\
3&    32.67&   36.23&   33.66&   \bf 58.96&   \bf 55.75&   \bf 54.87&   60.25&   56.28&   54.27&   {25.74}   \\
\hline\hline
\end{tabular*}
\end{table}

Note that due to the ``$-10\log_{10}$" definition, high CRIB in dB
means high accuracy, and vice versa. A CRIB of 50 dB means that
the standard angular deviation (square root of mean square angular
error) of the factor is cca $0.18^{\text{o}}$; a CRIB of 20 dB
corresponds to the standard deviation $5.7^{\text{o}}$.

The second mode to the decomposition, which represents intensity
of the data versus the emission wavelength, for $R=2,3,4$ and 8 is
shown in Figure 1. We can see that the CRIB allows to distinguish
between strong/significant modes of the decomposition and possibly
artificial modes due to over-fitting the model. The criterion is
different in general than the plain ``energy" of the factor; if a
factor has a low energy, it will probably have high CRIB, but it
might not hold true vice versa. A high energy component might have
a high CRIB.

In the next experiment, we have studied how much the accuracy of
the decomposition is affected in case that some data are missing
(not available). The decomposition with the correct rank $R=3$ and
$\sigma^2$ estimated as in (\ref{estsig}) was taken as a ground
truth; the 0-1 indicator tensor $\tW$ of the same size was
randomly generated with a given percentage of missing values. The
CRIB of the second mode factors was plotted in Figure 2 as a
function of this missing value rate. The figure also contains mean
square angular error of the components obtained in simulations.
Here an artificial Gaussian noise with zero mean and variance
$\sigma^2$ was added to the ``ground truth" tensor. The
decomposition was obtained by a Levenberg-Marquardt algorithm
\cite{SIAM} modified for tensors with missing entries.

A few observations can be made here.
\begin{itemize}
\item CRIB coincides with MSAE for the percentage of the missing
entries smaller than 70\%. If the percentage exceeds the
threshold, CRIB becomes overly optimistic. \item In general,
accuracy of the decomposition declines slowly with the number of
missing entries. If the number of missing entries is about 20\%,
loss of accuracy of the decomposition is only about 1-2 dB.
\end{itemize}

\begin{figure}
\centering
\subfigure[Estimated components as $R$ = 2.]{
\includegraphics[width=.46\linewidth, trim = 0.2cm 0cm 0cm 0.2cm,clip=true]{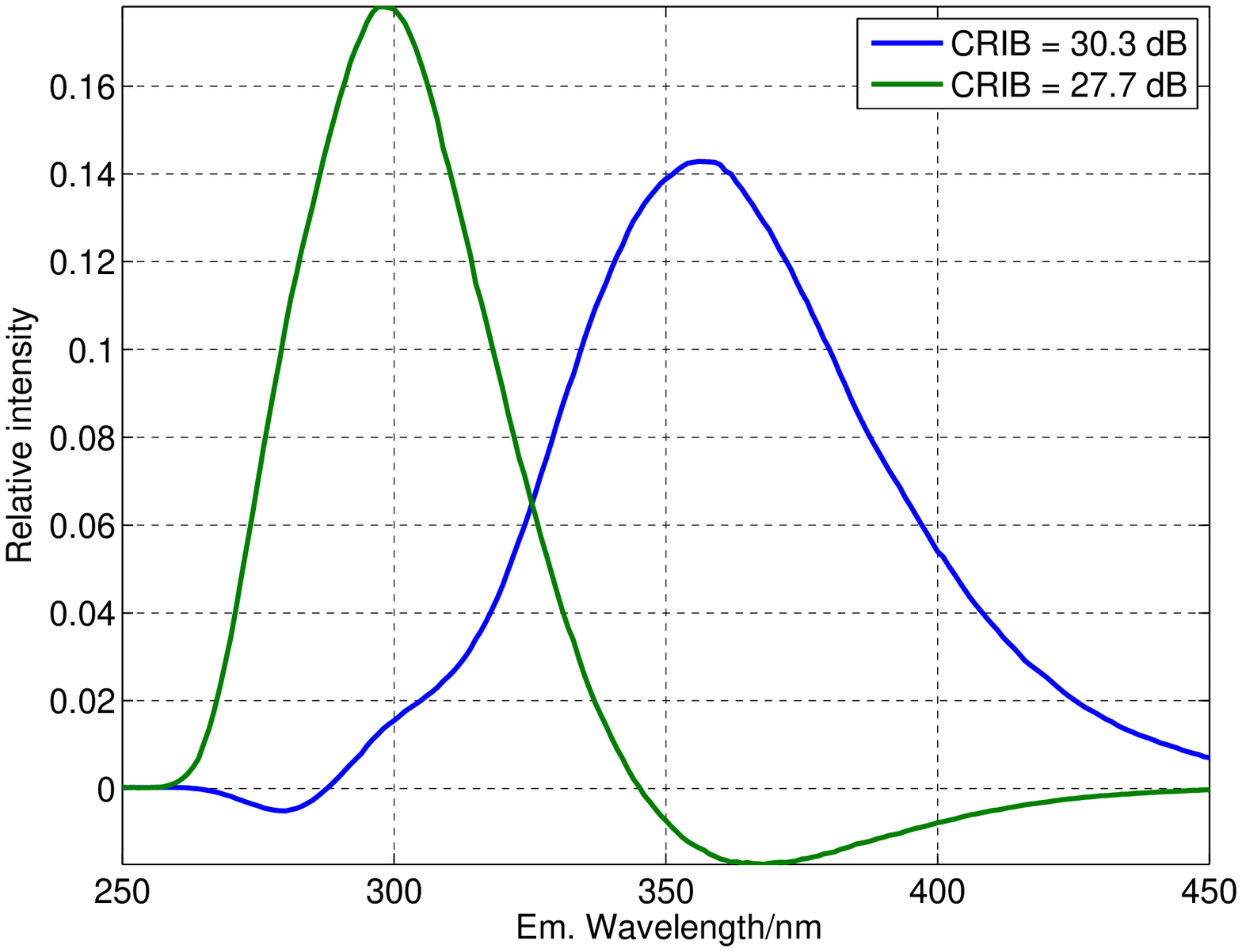}\label{fig_amino_CP_2vsCRIB}
} \hfill \subfigure[Estimated components as $R$ = 3.]{
\includegraphics[width=.46\linewidth, trim = 0.2cm 0cm 0cm 0.2cm,clip=true]{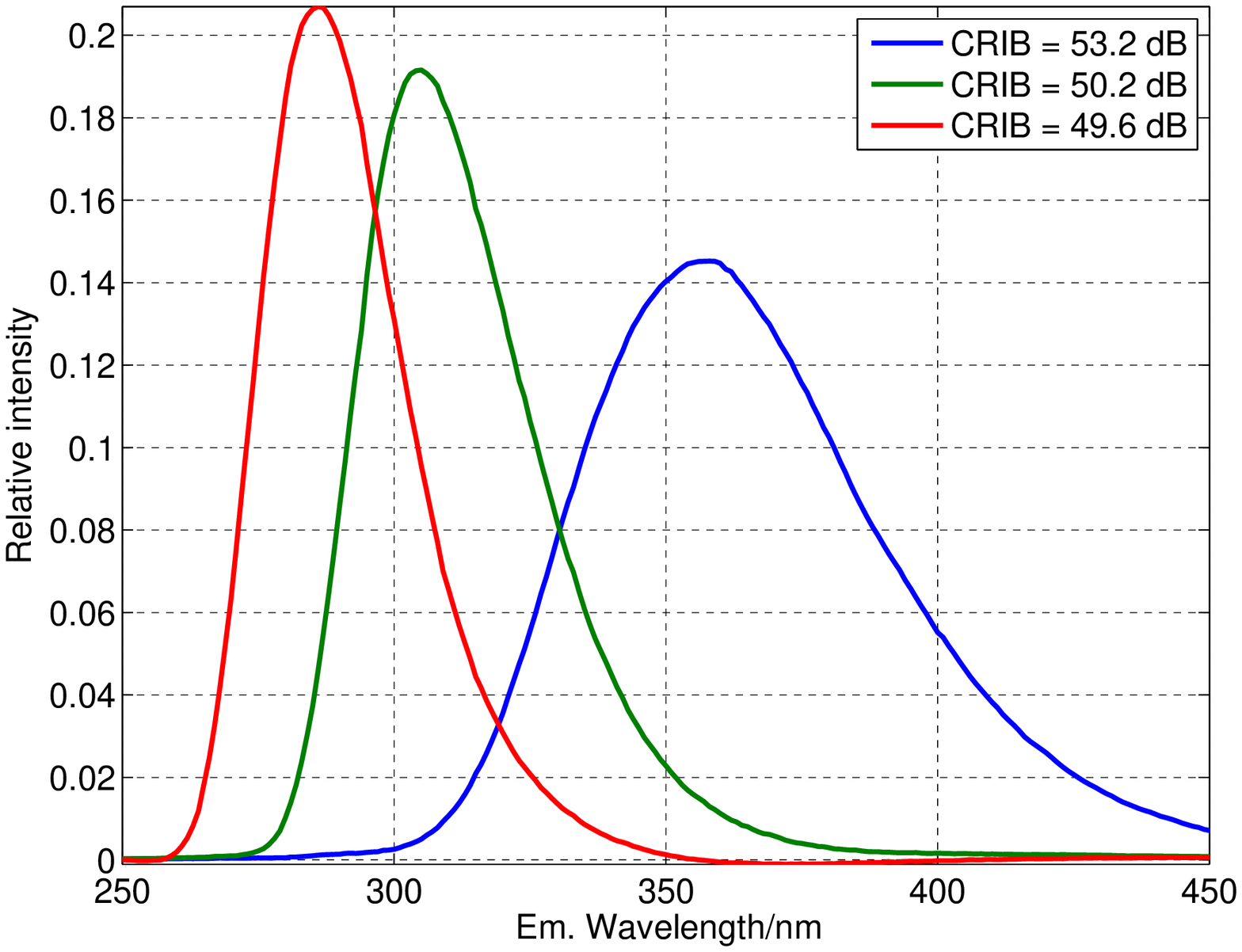}\label{fig_amino_CP_3vsCRIB}}
\subfigure[Estimated components as $R$ = 4.]{
\includegraphics[width=.46\linewidth, trim = 0.2cm 0cm 0cm 0.2cm,clip=true]{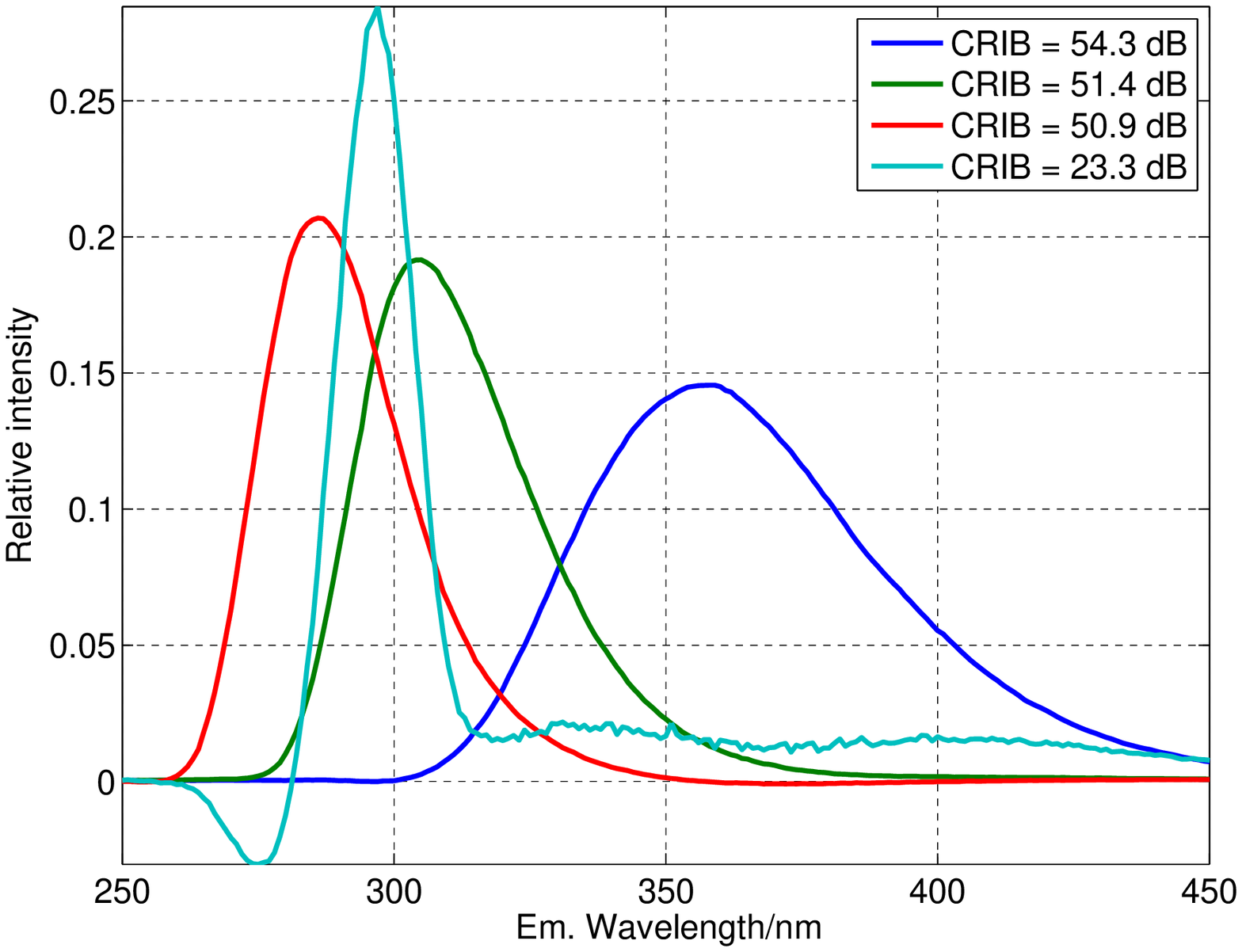}\label{fig_amino_CP_4vsCRIB}
}
\hfill
 \subfigure[Estimated components as $R$ = 8.]{
\includegraphics[width=.48\linewidth, trim = 0.2cm 0cm 1.3cm 1.0cm,clip=true]{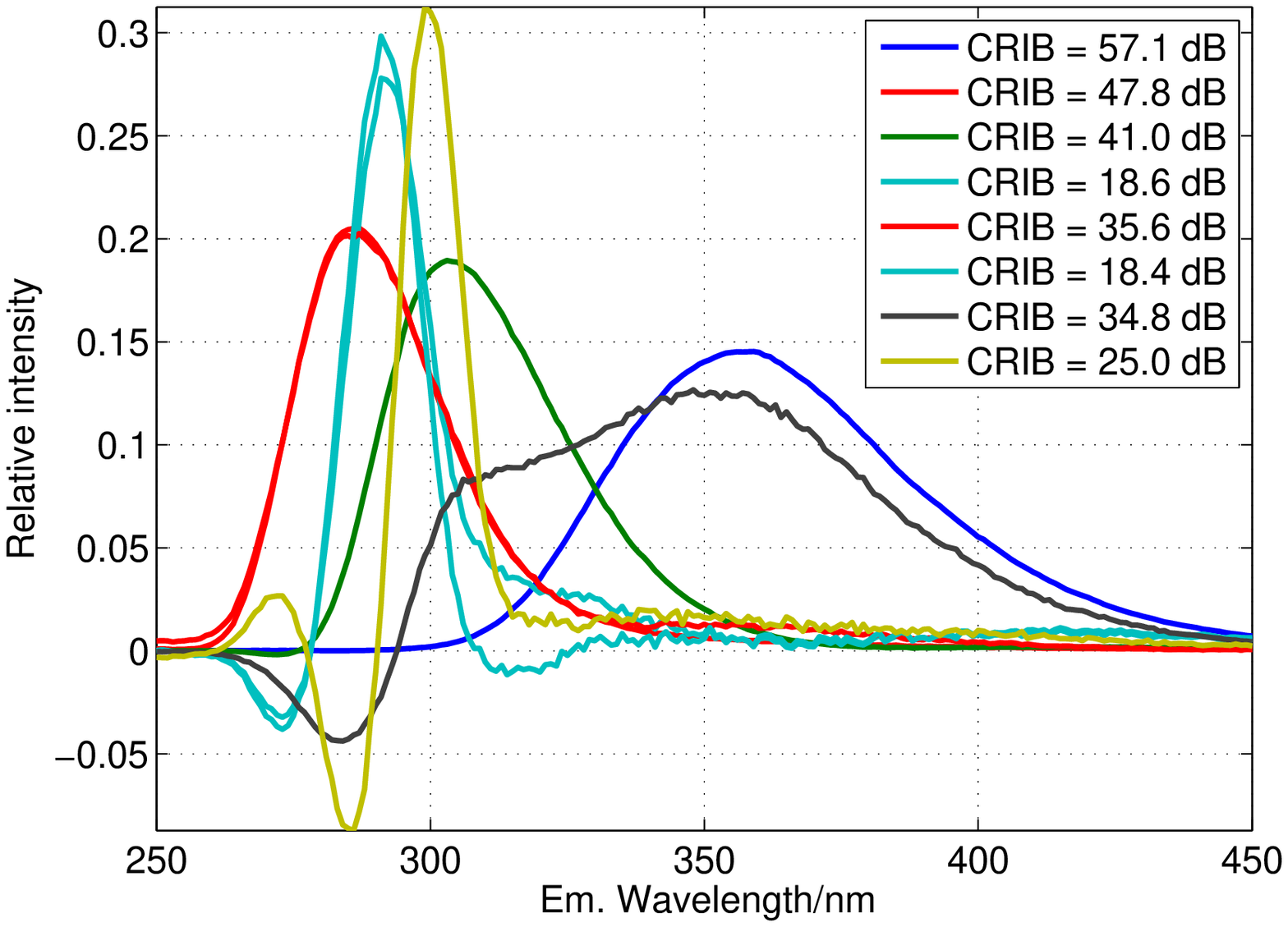}\label{fig_amino_CP_8vsCRIB}
} \caption{Illustration for emission components from best-fit
decompositions over 100 Monte Carlo runs for example VI-A. }
\label{fig_amino_RvsCRIB}
\end{figure}

\begin{figure}
\includegraphics[width=\linewidth]{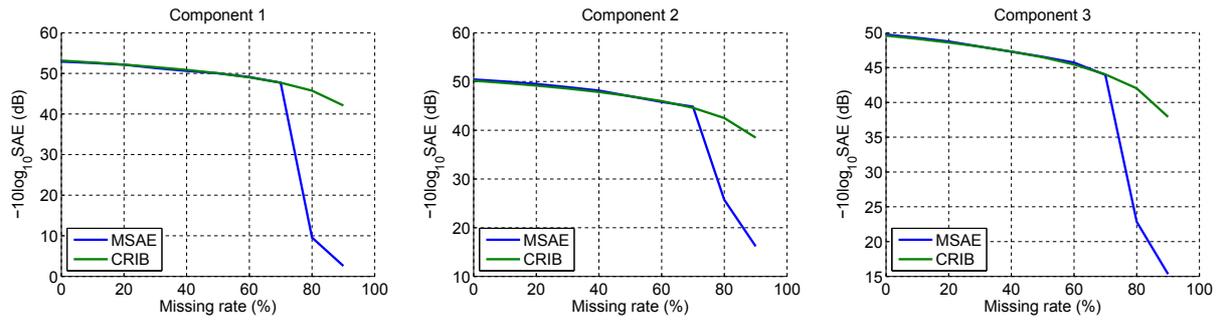} 
\caption{CRIB for the second-mode components of CP decomposition
of tensor in section VI.A with missing elements and mean square
angular error obtained in simulations versus percentage of the
missing elements.} \label{fig_amino_CP_CRIBvsMSAE}
\end{figure}

\subsection{Stability of the decomposition of Brie's tensor}

Brie {\em et al} \cite{Brie} presented an example of a four-way
tensor of rank 3, which arises while studying the response of
bacterial bio-sensors to different environmental agents. The
tensor has co-linear columns in three of four modes and the main
message of the paper is that its CP decomposition is still unique.
In this subsection we verify stability of the decomposition.

The factor matrices of the tensor have the form
$$
 \bA_1=[\ba_1,\ba_2,\ba_3]\quad \bA_2=[\ba_4,\ba_4,\ba_5],\quad
 \bA_3=[\ba_6,\ba_7,\ba_6]\quad \bA_4=[\ba_8,\ba_9,\ba_9]~.
$$
Assume for simplicity that all factors have unit norm,
$\|\ba_n\|=1$, $n=1,\ldots,9$. Due to Theorem 5 it holds that CRIB
on $\ba_1$ is a function of scalars $c_{11}=\ba_1^T\ba_2$,
$c_{12}=\ba_1^T\ba_3$, $c_{13}=\ba_2^T\ba_3$, $c_2=\ba_4^T\ba_5$,
$c_3=\ba_6^T\ba_7$, $c_4=\ba_8^T\ba_9$ and $I_1$, which is the
dimension of $\ba_1$. Then, the matrices $\bC_n=\bA_n^T\bA_n$,
$n=2,3,4$, have the form
$$
 \bC_2=\left[\begin{array}{ccc}1 & 1 & c_2\\ 1 & 1 & c_2\\ c_2 & c_2 &
 1\end{array}\right],
 \bC_3=\left[\begin{array}{ccc}1 & c_3 & 1\\ c_3 & 1 & c_3\\ 1 & c_3 & 1\end{array}\right],
 \bC_4=\left[\begin{array}{ccc}1 & c_4 & c_4\\ c_4 & 1 & 1\\ c_4 & 1 &
 1\end{array}\right]~.
$$
A straightforward usage of Theorem 4 is not possible, because some
of the involved matrices become singular. The CRIB itself,
however, is finite and can be computed using an artificial
parameter $\varepsilon$ as a limit. The limit CRIB is computed for
modified matrices at $\varepsilon\rightarrow 0$,
$$
 \bC_{2\varepsilon}=\left[\begin{array}{ccc}1 & 1-\varepsilon & c_2\\ 1-\varepsilon & 1 & c_2\\ c_2 & c_2 &
 1\end{array}\right],
 \bC_{3\varepsilon}=\left[\begin{array}{ccc}1 & c_3 & 1-\varepsilon\\ c_3 & 1 & c_3\\ 1-\varepsilon & c_3 & 1\end{array}\right],
 \bC_{4\varepsilon}=\left[\begin{array}{ccc}1 & c_4 & c_4\\ c_4 & 1 & 1-\varepsilon\\ c_4 & 1-\varepsilon &
 1\end{array}\right]~.
$$
If any of the correlations $c_2,c_3,c_4$ is zero, it is also
augmented by $\varepsilon$.

The limit CRIB can be shown to be independent of off-diagonal
elements of $\bC_1$, unless $\bC_1$ is singular. Assume that
$\bC_1$ is regular.
The result, obtained by Symbolic Matlab, is
\begin{eqnarray}
\mbox{CRIB}_{\varepsilon=0}(\ba_1)&=&
\frac{\sigma^2}{\|\ba_1\|^2}\,\frac{1}{2c_2^2c_3^2c_4^2 -
c_2^2c_3^2 - c_2^2c_4^2 - c_3^2c_4^2 +
1}\Biggl[(I_1-1)(1-c_2^2c_3^2)\nonumber\\&& -
\frac{c_3^4(c_2^2+1)-3c_3^2+1}{1-c_3^2}-\frac{c_2^4(c_3^2+1)-3c_2^2+1}{1-c_2^2}
+\frac{2-c_2^2-c_3^2}{1-c_4^2}\Biggr]~.
\end{eqnarray}
It follows that the decomposition is stable, unless all three
factors in some mode are collinear.

\subsection{Maximum Stable Rank}

A theoretically interesting question is, what is the maximum rank
of a tensor of a given dimension which has a stable CP
decomposition (with finite CRIB). For easy reference, we shall
call it {\em maximum stable rank} and denote it
$R_{smax}(I_1,\ldots,I_N)$.

An upper bound for the maximum stable rank can be deduced from the
requirement that the number of free parameters in the model, which
is $R(\sum_{n=1}^N I_n - N+1)$ in CP decomposition, cannot exceed
dimension of the available data, which is $\prod_{n=1}^N I_n$. It
follows that
\begin{eqnarray}
R_{smax}(I_1,\ldots,I_N)\leq \left\lfloor\frac{\prod_{n=1}^N
I_n}{\sum_{n=1}^N I_n - N+1}\right\rfloor\label{smax}
\end{eqnarray}
where $\lfloor x\rfloor$ denotes the lower integer part of $x$. It
can be verified numerically that for many (and maybe
all\footnote{We do not have yet a formal proof that the equality
in (\ref{smax}) holds for {\em all} tensor dimensions and
orders.}) tensor dimensions, an equality in (\ref{smax}) holds. In
other words, it means that the CRIB computed, e.g., via Theorem 4
for a CP decomposition with rank $R=R_{smax}$ and some (e.g.
random) factor matrices is finite. For example, the maximum stable
rank is $R_{smax}=2$ for $2\times 2\times 2$ tensors, and
$R_{smax}=3$ for $3\times 3\times 3$ tensors. For order-8 tensors
of dimension $2\times\ldots\times 2$, $(8\times)$, it holds
$R_{smax}=28$.

It might be interesting to compare the maximum stable rank with
the maximum rank and the maximum typical rank (to be explained
below) for given tensor dimension, if they are known
\cite{typical}. If the elements of a tensor are chosen randomly
according to a continuous probability distribution, there is not a
rank which occurs with probability 1 in general. Such rank, if
exists, is called generic. Ranks which occur with strictly
positive probabilities are called typical ranks. For example it
was computed in \cite{222} that probability for a real random
Gaussian tensor of the size $2\times 2\times 2$ to be 2 and 3 is
$\pi/4$, and $1-\pi/4$, respectively. We can see that no tensor of
the rank 3 and the dimension has a stable decomposition. For
tensors of the dimension $3\times 3\times 3$ the typical rank is 5
\cite{222}, it is a generic rank - but no decomposition of these
rank-5 tensors is stable, as $R_{smax}=3$.

Next, it might be interesting to compare the maximum stable rank
with the maximum rank for unique tensor decomposition, or prove
that these two coincide. Liu and Sidiropoulos
\cite{NWAYunique,CRB} derived a necessary condition for uniqueness
of the CP decomposition, which, according to a formulation in
\cite{Kolda} reads
\begin{eqnarray}
\min_{n=1,\ldots,N} \mbox{rank}\left(\bA_1\odot \ldots
\odot\bA_{n-1}\odot \bA_{n+1}\odot\ldots\odot \bA_N\right) =
R\label{cond0}
\end{eqnarray}
where $\odot$ means the Khatri-Rao product. The condition
(\ref{cond0}) is equivalent to the condition that the matrices
$\Zetab_n=\bA_1\odot \ldots \odot\bA_{n-1}\odot
\bA_{n+1}\odot\ldots\odot \bA_N$ have all full column rank,
$n=1,\ldots,N$, which is further equivalent to the condition that
the product $\Zetab_n^T\Zetab_n$ are regular for $n=1,\ldots,N$.
Finally note that
$$
\Zetab_n^T\Zetab_n=\bGamma_{nn},\qquad n=1,\ldots,N~.
$$
where $\bGamma_{nn}$ was defined in (\ref{defC}) and appears in
computation of the CRIB.

Unfortunately, it appears that the condition (\ref{cond0}) is only
necessary, but not sufficient for uniqueness. It is often
fulfilled for $R$ higher than $R_{smax}$. Thus a relation between
the stability and uniqueness of the CP decomposition remains open
question for now.

\section{Conclusions}
Cram\'er-Rao bounds for CP tensor decomposition represent an
important tool for studying
 accuracy and stability of the decomposition. The bounds
derived in this manuscript serve as a theoretical support for a
method of the decomposition through tensor reshaping \cite{3w2Nw}.
As a side result, a novel method of inverting Hessian matrix,
which is more computationally efficient, is derived for the
problem. It enables a further improvement of speed of the fast
Gauss-Newton for the problem \cite{SIAM}. A novel expression for
Hessian for CP decomposition of tensor with missing entries has
been derived. It can serve for assessing accuracy of CP
decomposition of these tensors without need of long Monte Carlo
simulations, and for implementing a damped Gauss-Newton algorithm
for CP decomposition of these tensors.

A direct link between stability and essential uniqueness remains
to be an open theoretical question. In particular, it is not known
yet for sure if stability implies the essential uniqueness.

CRB expressions similar to the ones derived in this paper can be
also derived for other important special tensor decomposition
models such as INDSCAL (where two or more factor matrices
coincide) \cite{stegeman,Tendeiro}, or for the PARALIND model,
where the factor matrices have certain structure
\cite{stegeman12b}, and for block factorization methods.

\section*{Appendix A}
\noindent{\bf Matrix Inversion Lemma} (Woodbury identity)\\
Let $\bA$, $\bX$, $\bY$, and $\bR$ are matrices of compatible
dimensions such that the following products and inverses exist.
Then
\begin{eqnarray}
(\bA+\bX\bR\bY)^{-1}=\bA^{-1}-\bA^{-1}\bX(\bR^{-1}+\bY\bA^{-1}\bX)^{-1}\bY\bA^{-1}~.\label{MIL}
\end{eqnarray}

\section*{Appendix B}
\noindent{\bf Proof of Theorem 4}\\
Let the matrices $\bK$ and $\bPsi$ in (\ref{defB}) be partitioned
as
\begin{eqnarray}
\bK=\left[\begin{array}{cc}{\bf 0} & \bK_1\\
\bK_1^T & \bK_2\end{array}\right],\qquad
\bPsi=\left[\begin{array}{cc}\bPsi_1 & {\bf 0}\\
{\bf 0} & \bPsi_2\end{array}\right]
\end{eqnarray}
where the left-upper blocks have the size $R^2\times R^2$. Then,
using a formula for inverse of partitioned matrices, the
left-upper block of $\bB$ in (\ref{defB}) can be written as
\begin{eqnarray}
\bB_0=\bK_1(\bI_{(N-1)R^2}+\bPsi_2\bK_2-\bPsi_2\bK_1^T\bPsi_1\bK_1)^{-1}\bPsi_2\bK_1^T\stackrel{\triangle}{=}
\bK_1\bK_3^{-1}\bPsi_2\bK_1^T~.\label{defK3}
\end{eqnarray}
A key observation which enables a fast inversion of the term
$\bK_3$ is that
\begin{eqnarray}
\bK&=&\bK_0+\bD\bF\bD^T
\end{eqnarray} where
\begin{eqnarray}\bK_0&=&-{\tt{bdiag}}\left(\bP_R\bF({\tt dvec}(1\oslash \bC_n))^2\right)_{n=1}^N\\
\bF&=&\bP_R\,\prod_{n=1}^N {\tt dvec}(\bC_n)=\bP_R\,{\tt dvec}(\bGamma_{11}\circledast\bC_1)\\
\bD&=&\left[{\tt dvec}(1\oslash \bC_1),\ldots,{\tt dvec}(1\oslash
\bC_N)\right]^T~.
\end{eqnarray}
Similarly,
\begin{eqnarray}
\bK_2&=&\bK_{02}+\bD_2\bF\bD_2^T
\end{eqnarray} where
\begin{eqnarray}\bK_{02}&=&-{\tt{bdiag}}\left(\bP_R\bF({\tt dvec}(1\oslash \bC_n))^2\right)_{n=2}^N\\
\bD_2&=&\left[{\tt dvec}(1\oslash \bC_2),\ldots,{\tt
dvec}(1\oslash \bC_N)\right]^T~.
\end{eqnarray}
Then the matrix $\bK_3$ in (\ref{defK3}) can be written as
\begin{eqnarray}
\bK_3&=&
\bI_{(N-1)R^2}+\bPsi_2\bK_2-\bPsi_2\bK_1^T\bPsi_1\bK_1\nonumber\\
&=&
\bI_{(N-1)R^2}+\bPsi_2(\bK_{02}-\bK_1^T\bPsi_1\bK_1)+\bPsi_2\bD_2\bF\bD_2^T\nonumber\\
&=& \bQ+\bPsi_2\bD_2\bF\bD_2^T
\end{eqnarray}
where
\begin{eqnarray}
\bQ&=&{\tt{bdiag}}(\bQ_n)_{n=2}^N\\
\bQ_n&=& \bI_{R^2}-(\bGamma_{nn}^{-1} \otimes
\bX_n)\bP_R\,(\bF({\tt dvec}(1\oslash \bC_n))^2+{\tt
dvec}(\bGamma_{1n})(\bGamma_{11}^{-1} \otimes \bC_1){\tt
dvec}(\bGamma_{1n})\bP_R)~.
\end{eqnarray}
Now, $\bK_3$ can be easily inverted using the matrix inversion
lemma (\ref{MIL}),
\begin{eqnarray}
\bK_3^{-1} &=& \bQ^{-1}-\bQ^{-1}\bD_2^T(\bI_{R^2}+\bD_2^T
 \bQ^{-1}\bPsi_2\bD_2\bF)^{-1}\bPsi_2\bD_2\bF\bQ^{-1}~.\label{comK3}
\end{eqnarray}
Inserting (\ref{comK3}) in (\ref{defK3}) gives, after some
simplifications, the result (\ref{result4}).\hfill\rule{2mm}{2mm}

\section*{Appendix C}
\noindent{\bf Proof of Theorem 5}\\
Consider the change of scale of columns of factor matrices up to
their first columns. As in Section II assume that the scale change
is realized in $\bA_1$, while the other factor matrices have
columns of unit norm. The theorem claims that the substitution
$\bA_1\leftarrow\bA_1\bD$ into (\ref{result}) where
$\bD={\tt{diag}}(1,\lambda_2,\dots,\lambda_R)$, $\lambda_r\neq 0$,
has no influence on $\mbox{CRIB}(\ba_1)$.

The substitution $\bA_1\leftarrow\bA_1\bD$ leads to
$\bC_1\leftarrow\bD\bC_1\bD$ and $\bX_1\leftarrow\bD\bX_1\bD$
while $\bC_n$ and $\bX_n$, $n=2,\dots,N$, remain the same.
Consequently, $\bGamma_{1n}$, $n=1,\dots,N$, remain unchanged
while $\bGamma_{nn}\leftarrow\bD\bGamma_{nn}\bD$ for
$n=2,\dots,N$. Now, we can substitute into (\ref{result4})
assuming that the condition of Theorem~4 is satisfied.

Let $\widetilde{\bS}_n$ denote the matrix $\bS_n$ in (\ref{defSn})
after the substitution $\bA_1\leftarrow\bA_1\bD$. It can be shown
that $(\bD\otimes\bI_R)\widetilde{\bS}_n=\bS_n(\bD\otimes\bI_R)$
using the rules
\begin{align} (\bD\bGamma_{nn}\bD)^{-1}\otimes\bX_n&=(\bD^{-1}\otimes\bI_R)(\bGamma_{nn}^{-1}\otimes\bX_n)(\bD^{-1}\otimes\bI_R)\\
{\tt dvec}(\bD\bGamma_{nn}\bD\oslash\bC_n)&=(\bD\otimes\bD){\tt dvec}(\bGamma_{nn}\oslash\bC_n)\\
(\bI_R\otimes\bD)\bP_R&=\bP_R(\bD\otimes\bI_R)
\end{align}
and the fact that diagonal matrices commute. Using the same rules
in further substitutions, after some computations, the
independence of $\mbox{CRIB}(\ba_1)$ on $\bD$ follows.

\section*{Appendix D}
\noindent{\bf Proof of Theorem 6}\\
Again, assume for simplicity that all factors have unit norms. It
holds
$$
\bGamma_{11}=\left[\begin{array}{cc} 1 & h_1\\
h_1 & 1\end{array}\right],\quad \bX_n=\left[\begin{array}{cc} 0 & 0\\
0 & 1-c_n^2\end{array}\right],\quad n=1,\ldots,N~.
$$
and
\begin{eqnarray}
g_{11}&=&[\bGamma_{11}^{-1}]_{11}=\frac{1}{1 - h_1^2}\\
\bg_{1,:}&=&g_{11}\left[1,\quad - h_1\right]~.
\end{eqnarray}
The matrix $\bPsi$ in (\ref{defPsi}) can be decomposed as
$\bPsi=\bJ\bPhi$ where
\begin{eqnarray}
\bJ &=& {\tt{bdiag}}\left(\bI_4,\bI_2\otimes [0,\,
1]^T,\ldots,\bI_2\otimes [0 ,\,1]^T\right)\\
\bPhi &=& {\tt{bdiag}}\left(\bGamma_{11}^{-1}\otimes \bC_1,
(1-c_2^2)\bGamma_{22}^{-1}\otimes [0,\, 1],\ldots,
(1-c_N^2)\bGamma_{NN}^{-1}\otimes [0,\, 1]\right)~.
\end{eqnarray}
Then the matrix $\bB$ in (\ref{defB}) can be rewritten using the
Woodbury identity (\ref{MIL}) as
\begin{eqnarray}
\bB &=&
\bK(\I_{4N}+\bJ\bPhi\bK)^{-1}=\bK-\bK\bJ(\bI_{2N+2}+\bPhi\bK\bJ)^{-1}\bPhi\bK~.
\end{eqnarray}
Now, put $\bB_4=\bI_{2N+2}+\bPhi\bK\bJ$ and write it in the block
form as
\begin{eqnarray}
\bB_4 &=& \bI_{2N+2}+\bPhi\bK\bJ = \left[\begin{array}{cc}\bB_{41}
& \bB_{42}\\ \bB_{43} & \bB_{44}\end{array}\right]
\end{eqnarray}
where $\bB_{41}$ has the size $4\times 4$. The bottom-right block
$\bB_{44}$ of dimension $(2N-2)\times(2N-2)$ is easy to be
inverted using the Woodbury identity again, because it can be
written as
\begin{eqnarray}
\bB_{44} &=& \bB_5+\bs\bff^T
\end{eqnarray}
where
\begin{eqnarray}
\bB_5 &=& {\tt{bdiag}}\left(\bB_{52},\ldots,\bB_{5N}\right)\\
\bB_{5n}&=& \left[\begin{array}{cc}1 & -
\frac{h_nc_1(1-c_n^2)}{1-h_n^2c_1^2}\\ 0 &
\frac{c_n^2-h_n^2c_1^2}{1-h_n^2c_1^2}
\end{array}\right],\qquad n=2,\ldots,N\\
\bs&=& \left[- \frac{h_2c_1(1-c_2^2)}{1-h_2^2c_1^2},
\frac{(1-c_2^2)}{1-h_2^2c_1^2},\ldots,-
\frac{h_Nc_1(1-c_N^2)}{1-h_N^2c_1^2},
\frac{(1-c_N^2)}{1-h_N^2c_1^2}
\right]^T\\
\bff&=&[0,\,1,\,0,\,1,\ldots,1]^T~.
\end{eqnarray}
 After some computations, we receive the result
 (\ref{result31}).\hfill\rule{2mm}{2mm}

\section*{Appendix E}
\noindent{\bf Proof of Theorem 7}\\

Under the assumption of the Theorem, it holds that the matrix
$\bC_1=\bA_1^T\bA_1$ is diagonal and $\bC_2=\bI_R$ (identity
matrix). Thanks to Theorem 5 we can assume, without any loss of
generality, that $\bC_1=\bI_R$ as well. It can be shown for
$\bGamma_{mn}$ in (\ref{defC}) that $\bGamma_{mn}=\bI_R$ for all
pairs $(m,n)$, $(m,n)\ne (1,2),(2,1)$. Only $\bGamma_{12}$ and
$\bGamma_{21}=\bGamma_{12}$ are possibly different. Note that the
first row of $\bGamma_{12}$ is $(1,\gamma_2,\ldots,\gamma_N)$.

It follows from these observations that all non-diagonal
$R^2\times R^2$ blocks $\bK_{mn}$ of $\bK$ in (\ref{hessi}) with
$(m,n)\ne (1,2),(2,1)$ are identical, diagonal, having 1 at
positions $(p,p)$, $p=1,R+1,2R+2,\ldots,R^2$ and 0 elsewhere. In
other words, these $\bK_{mn}$ can be written as
$\bK_{mn}=\bQ\bQ^T$, where $\bQ$ is a 0-1 matrix of the size
$R^2\times R$, the $p-$th column of $\bQ$ has the value 1 at
position $(p-1)(R+1)+1$ and 0 elsewhere.

Computation of the CRIB can proceed from equation (\ref{defK3})
by inserting the special form
of the blocks of $\bK_1$ and $\bK_2$ and using the Woodbury identity
(\ref{MIL}). \hfill\rule{2mm}{2mm}

\section*{Appendix F}
\noindent{\bf Proof of Theorem 8}\\

The following identities are used in this proof
\begin{eqnarray}
    \vtr{\bA \circledast \bB} &=& \dvec(\bB) \, \vtr{\bA}, \\
    \ba^T  \, \diag(\bb) \, \bc &=& \left(\ba \circledast \bc \right)^T \bb, \\
    (\ba \otimes \bb) \circledast (\bc \otimes \bd) &=& (\ba \circledast \bc ) \otimes
    (\bb \circledast \bd )~.
\end{eqnarray}
Here, dimensions of $\ba$, $\bb$, $\bc$ and $\bd$ are assumed to
match accordingly.


The approximate Hessian in (\ref{51}) is given by
\begin{eqnarray}
    \bH = \bJ_W^T(\thetab) \bJ_W(\thetab) = \bJ(\thetab)^T \dvec(\tW) \bJ(\thetab),
\end{eqnarray}
where $\bJ(\thetab)$ is the Jacobian for the complete data.

We have
\begin{eqnarray}
 \frac{\partial \vtr{\tY}}{\partial \ba^{(n)}_{ir}} &= &
              \left(\bigotimes_{k = n+1}^{N} \ba^{(k)}_r\right) \otimes \ve^{(n)}_{i}   \otimes \left(\bigotimes_{k = 1}^{n-1} \ba^{(k)}_r\right)
           \, , \label{equ_grad_y_ar}
\end{eqnarray}
where unit vector $\ve^{(n)}_{i}$ for $i =1 , 2, \ldots, I_n$ is
the $i$-th column of the identity matrix  of size ${I_n \times
I_n}$.

An $(i,j)$ entry of a sub matrix $\bH^{(n,n)}_{r,s}$ for $i = 1,
2,\ldots, I_n$, and $j = 1, 2,\ldots, I_n$ is given by
\begin{eqnarray}
     \bH^{(n,n)}_{r,s}(i,j) &=& \left(\frac{\partial \vtr{\tY}}{\partial \ba^{(n)}_{ir}}   \right)^T \, \dvec{\left(\tW\right)} \;
\left( \frac{\partial \vtr{\tY}}{\partial \ba^{(n)}_{js}} \right)
  \notag\\
    &=&  \left( \left(\bigotimes_{k = n+1}^{N} \ba^{(k)}_r \circledast \ba^{(k)}_s\right) \otimes \left(\ve^{(n)}_{i} \circledast \ve^{(n)}_{j}\right)
    \otimes \left(\bigotimes_{k = 1}^{n-1} \ba^{(k)}_r \circledast \ba^{(k)}_s\right)  \right)^T  \; \vtr{\tW}
    \notag \\
    &=&  \tW \; {\bar\times}_{-n} \{\bb^{(k)}\}  \; {\bar\times}_n \, \delta_{ij} \, \ve^{(n)}_{i}
     \,, \label{equ_H_nmrs}
\end{eqnarray}
where $\delta_{ij}$ is the Kronecker delta, $\bb^{(n)} =
\ba^{(n)}_r \circledast \ba^{(n)}_s$, for $n = 1, \ldots, N$. This
leads to that a diagonal sub-matrix $\bH^{(n,n)}_{r,s}$ is a
diagonal matrix as in Theorem~\ref{theo_Hessian_missing}.

For off-diagonal sub matrices $\bH^{(n,m)}_{r,s}$ of size $I_n
\times I_m$ ($ 1\le n < m \le N$), we have
\begin{eqnarray}
     \bH^{(n,m)}_{r,s}(i,j) &=& \left(\frac{\partial \vtr{\tY}}{\partial \ba^{(n)}_{ir}}   \right)^T \, \dvec{\left(\tW\right)} \;
\left( \frac{\partial \vtr{\tY}}{\partial \ba^{(m)}_{js}} \right)
  \notag\\
    &=&  \left( \left(\bigotimes_{k = m+1}^{N} \ba^{(k)}_r \circledast \ba^{(k)}_s\right) \otimes \left(\ba^{(m)}_{r} \circledast \ve^{(m)}_{j}\right)
    \otimes
    \left(\bigotimes_{k = n+1}^{m-1} \ba^{(k)}_r \circledast \ba^{(k)}_s\right)     \otimes  \right.
    \notag\\
    &&\left.
    \left(\ve^{(n)}_{i} \circledast \ba^{(n)}_{s}\right)
    \otimes \left(\bigotimes_{k = 1}^{n-1} \ba^{(k)}_r \circledast \ba^{(k)}_s\right)  \right)^T  \; \vtr{\tW} \notag\\
    &= & a^{(m)}_{jr} \, a^{(n)}_{is} \, \left( \tW {\bar\times}_{-\{n,m\}} \, \{\bb^{(k)}\}
     \;  {\bar\times}_{n}  \ve^{(n)}_{i} \;  {\bar\times}_{m}  \ve^{(m)}_{j}\ \right).
\end{eqnarray}
This leads to the compact form in Theorem~8.\hfill\rule{2mm}{2mm}

\end{document}